\documentclass[traditabstract]{aa}

\usepackage{txfonts}

\usepackage{natbib}
\usepackage{graphicx}

\begin{document}
\title{WASP-78b and WASP-79b: Two highly-bloated hot Jupiter-mass
exoplanets orbiting F-type stars in Eridanus
\thanks{
Radial velocity and photometric data are only available in electronic form at the
CDS via anonymous ftp to\newline
{\tt cdsarc.u-strasbg.fr (130.79.128.5)} or via\newline
http://cdsarc.u-strasbg.fr/viz-bin/qcat?J/A+A/???/A??}}
\author
{B.~Smalley\inst{1},
D.~R.~Anderson\inst{1},
A.~Collier-Cameron\inst{2},
A.~P.~Doyle\inst{1},
A.~Fumel\inst{3},
M.~Gillon\inst{3},
C.~Hellier,\inst{1},
E.~Jehin\inst{2},
M.~Lendl\inst{4},
P.~F.~L.~Maxted\inst{1},
F.~Pepe\inst{4},
D.~Pollacco\inst{5},
D.~Queloz\inst{4},
D.~S\'egransan\inst{4},
A.~M.~S.~Smith\inst{1},
J.~Southworth\inst{1}
A.~H.~M.~J.~Triaud\inst{4},
S.~Udry\inst{4},
R.~G.~West\inst{6},
}

\authorrunning{B. Smalley et al.}
\titlerunning{WASP-78b and WASP-79b}

\institute{
Astrophysics Group, Keele University, Staffordshire, ST5 5BG, United Kingdom\newline
\email{b.smalley@keele.ac.uk}
\and
SUPA, School of Physics and Astronomy, University of St. Andrews, North Haugh, Fife, KY16 9SS, United Kingdom
\and
Universit\'e de Li\`ege, All\'ee du 6 ao\^ut 17, Sart Tilman, Li\`ege 1, Belgium
\and
Observatoire de Gen\`eve, Universit\'e de Gen\`eve, Chemin des maillettes 51, 1290 Sauverny, Switzerland
\and
Astrophysics Research Centre, School of Mathematics \& Physics, Queen's University Belfast, University Road, Belfast BT7 1NN, United Kingdom
\and
Department of Physics and Astronomy, University of Leicester, Leicester, LE1 7RH, United Kingdom
}

\date{Received <date> / accepted <date>}

\abstract{We report the discovery of WASP-78b and WASP-79b, two highly-bloated
Jupiter-mass exoplanets orbiting F-type host stars. WASP-78b orbits its $V=12.0$
host star (TYC 5889-271-1) every 2.175~days and WASP-79b orbits its $V=10.1$
host star (CD-30 1812) every 3.662~days. Planetary parameters have been
determined using a simultaneous fit to WASP and TRAPPIST transit photometry and
CORALIE radial-velocity measurements. For WASP-78b a planetary mass of 0.89
$\pm$ 0.08 $M_{\rm Jup}$ and a radius of 1.70 $\pm$ 0.11 $R_{\rm Jup}$ is found.
The planetary equilibrium temperature of $T_{\rm P} = 2350 \pm 80$~K for
WASP-78b makes it one of the hottest of the currently known exoplanets. WASP-79b
its found to have a planetary mass of 0.90 $\pm$ 0.08 $M_{\rm Jup}$, but with a
somewhat uncertain radius due to lack of sufficient TRAPPIST photometry. The
planetary radius is at least 1.70 $\pm$ 0.11 $R_{\rm Jup}$, but could be as
large as 2.09 $\pm$ 0.14 $R_{\rm Jup}$, which would make WASP-79b the largest
known exoplanet.}


\maketitle

\section{Introduction}

The first exoplanets were discovered using the radial velocity technique
\citep{1995Natur.378..355M}. However, following the detection of a transiting
exoplanet \citep{2000ApJ...529L..45C}, several ground-based and space-based
survey projects have dramatically increased the number of known systems.
Transiting exoplanets allow parameters such as the mass, radius, and density to
be precisely determined, as well as their atmospheric properties to be studied
during their transits and occultations
\citep{2005ApJ...626..523C,2009MNRAS.394..272S,2009IAUS..253...99W}.

Most of the transiting exoplanets found by ground-based surveys are 'hot
Jupiters', with orbital periods of up to around 5~days. Many of these have radii
larger than predicted by irradiated planet models \citep{2007ApJ...659.1661F}.
Several have markedly low densities, with WASP-17b
\citep{2010ApJ...709..159A,2011MNRAS.416.2108A}, Kepler-12b
\citep{2011ApJS..197....9F} and Kepler-7b \citep{2010ApJ...713L.140L} having a
density less than 1/10 that of Jupiter. The mechanisms for producing such
bloated planets are at present unclear \citep{2010SSRv..152..423F}, but several
have been proposed, including Ohmic heating in the planetary atmosphere
\citep{2010ApJ...714L.238B,2010ApJ...724..313P} and thermal tidal effects
\citep{2010ApJ...714....1A}.

In this paper we report the detection of WASP-78b and WASP-79b, two
highly-bloated Jupiter-mass planets in orbit around F-type stars. We present the
WASP-South discovery photometry, together with follow-up optical photometry and
radial velocity measurements.

\section{WASP-South photometry}
\label{WASP}

The WASP project has two robotic observatories; one on La Palma in the Canary
Islands and another in Sutherland in South Africa. The wide angle survey is
designed to find planets around relatively bright stars in the $V$-magnitude
range $9\sim13$. A detailed description is given in \citet{2006PASP..118.1407P}.

\begin{figure}[h]
\includegraphics[width=\columnwidth]{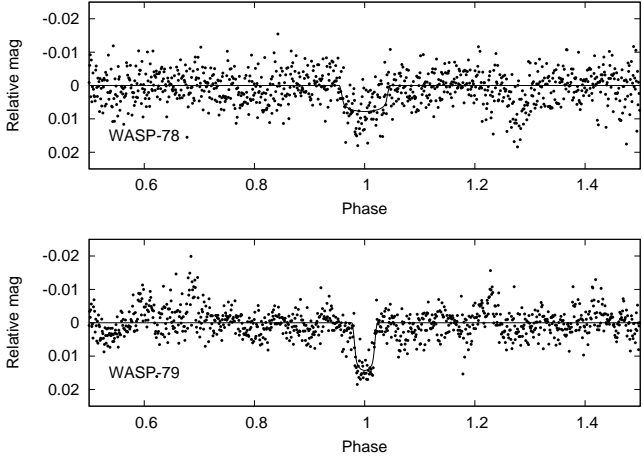}
\caption{WASP photometry of WASP-78 and WASP-79 folded on the best-fitting
orbital periods and binned into phase bins of 0.001. The solid line is the
model fit presented in Sect.~\ref{MCMC}.}
\label{WASP-phot}
\end{figure}

The pipeline-processed data were de-trended and searched for transits using the
methods described in \citet{2006MNRAS.373..799C}, yielding detections of
periodic, transit-like signatures on two stars in the constellation Eridanus
(Fig.~\ref{WASP-phot}). The $V=12.0$ star WASP-78 (1SWASPJ041501.50-220659.0;
TYC 5889-271-1) exhibited $\sim$0.010~{mag}. transits every 2.175~days, while
the $V=10.1$ star WASP-79 (1SWASPJ042529.01-303601.5; CD-30 1812) showed
$\sim$0.015~{mag.} transits every 3.66~days. A total of 16489 observations of
WASP-78 were obtained between 2006 August and 2009 December and 15424
observations of WASP-79 were obtained between 2006 September and 2010 February.

There is a 15$^{\rm th}$~{mag.} star, 2MASS\,04150416-2207189, located
42\arcsec\ away from WASP-78, which is just within the photometric extraction
aperture. However, the dilution is only 2\% and does not significantly affect
the depth of the WASP transits. Around 24\arcsec\ away from WASP-79, is
6dFGS\,gJ042530.8-303554, a 16$^{\rm th}$~{mag.} redshift $z = 0.069$ galaxy
\citep{2009MNRAS.399..683J}. This is, however, too faint to significantly dilute
the WASP-79 photometry. The spectral type of WASP-79 is listed as F2 in
\citet{1954AnCap..17.....J}.

\section{Spectroscopic observations with CORALIE}
\label{CORALIE_Spec}

Spectroscopic observations were obtained with the CORALIE spectrograph on the
Swiss 1.2m telescope. The data were processed using the standard pipeline
\citep{1996A&AS..119..373B,2000A&A...354...99Q,2002A&A...388..632P}. A total of
17 and 21 radial velocity (RV) and line bisector span ($V_{\rm span}$)
measurements were made for WASP-78 and WASP-79, from 2011 October 09 to 2011
December 30, and 2010 December 12 to 2012 February 07, respectively
(Table~\ref{rv-data}). The bisector spans are a measure of the asymmetry of the
cross-correlation function and, based on our experience, have standard errors of
$\approx2\sigma_{\rm RV}$.

\begin{table} 
\caption{Radial velocity (RV) and line bisector spans ($V_{\rm span}$)
measurements for WASP-78 and WASP-79 obtained by CORALIE spectra.} 
\label{rv-data} 
\centering
\begin{tabular}{lll} \hline\hline
BJD--2\,400\,000 (UTC)& RV (km\,s$^{-1}$) & $V_{\rm span}$ (km\,s$^{-1}$) \\ \hline
\multicolumn{3}{c}{WASP-78} \\
55843.80914 & 0.3565 $\pm$ 0.0349 & +0.0443  \\
55844.85276 & 0.6078 $\pm$ 0.0404 & +0.2036  \\
55856.82996 & 0.3477 $\pm$ 0.0388 &$-$0.0086 \\
55858.75524 & 0.3932 $\pm$ 0.0409 & +0.0772  \\
55859.84448 & 0.5282 $\pm$ 0.0339 & +0.1423  \\
55863.77620 & 0.4233 $\pm$ 0.0281 & +0.1274  \\
55864.77388 & 0.5390 $\pm$ 0.0284 & +0.1252  \\
55868.78606 & 0.5463 $\pm$ 0.0320 & +0.1018  \\
55872.82630 & 0.5131 $\pm$ 0.0346 &$-$0.1311 \\
55880.77994 & 0.3302 $\pm$ 0.0240 & +0.1064  \\
55881.82771 & 0.5734 $\pm$ 0.0272 & +0.0760  \\
55883.69133 & 0.5093 $\pm$ 0.0260 & +0.1505  \\
55887.76423 & 0.4160 $\pm$ 0.0231 & +0.0627  \\
55888.64129 & 0.5431 $\pm$ 0.0281 & +0.0722  \\
55889.62002 & 0.3780 $\pm$ 0.0257 & +0.0949  \\
55894.74373 & 0.5875 $\pm$ 0.0238 & +0.0426  \\
55925.58825 & 0.5346 $\pm$ 0.0206 & +0.0972  \\
\multicolumn{3}{c}{WASP-79} \\
55542.63132 & 4.8915 $\pm$ 0.0260 &$-$0.2869 \\
55573.65244 & 5.1141 $\pm$ 0.0225 &$-$0.1877 \\
55595.61945 & 5.0768 $\pm$ 0.0267 &$-$0.2985 \\
55832.81511 & 4.9912 $\pm$ 0.0280 &$-$0.3792 \\
55856.85501 & 4.9258 $\pm$ 0.0319 &$-$0.2776 \\
55858.83490 & 4.9823 $\pm$ 0.0315 &$-$0.1275 \\
55863.73949 & 4.9772 $\pm$ 0.0232 &$-$0.2613 \\
55864.82478 & 4.8688 $\pm$ 0.0236 &$-$0.2901 \\
55865.67114 & 4.9591 $\pm$ 0.0282 &$-$0.3785 \\
55867.86163 & 4.9566 $\pm$ 0.0241 &$-$0.2958 \\
55868.67269 & 4.9228 $\pm$ 0.0267 &$-$0.2961 \\
55869.70268 & 5.0298 $\pm$ 0.0266 &$-$0.2612 \\
55870.84257 & 5.0119 $\pm$ 0.0249 &$-$0.2504 \\
55871.79821 & 4.9102 $\pm$ 0.0249 &$-$0.3420 \\
55872.78874 & 4.9724 $\pm$ 0.0251 &$-$0.3896 \\
55873.83187 & 5.0848 $\pm$ 0.0260 &$-$0.3053 \\
55874.83089 & 5.1331 $\pm$ 0.0486 &$-$0.6088 \\
55883.76896 & 4.9481 $\pm$ 0.0252 &$-$0.2040 \\
55886.77560 & 4.9336 $\pm$ 0.0300 &$-$0.3774 \\
55888.67590 & 5.0633 $\pm$ 0.0258 &$-$0.1902 \\
55925.65354 & 5.0572 $\pm$ 0.0200 &$-$0.2109 \\
55964.62745 & 5.0409 $\pm$ 0.0216 &$-$0.3219 \\
\hline
\end{tabular} 
\end{table} 

\begin{figure}[h]
\includegraphics[width=\columnwidth]{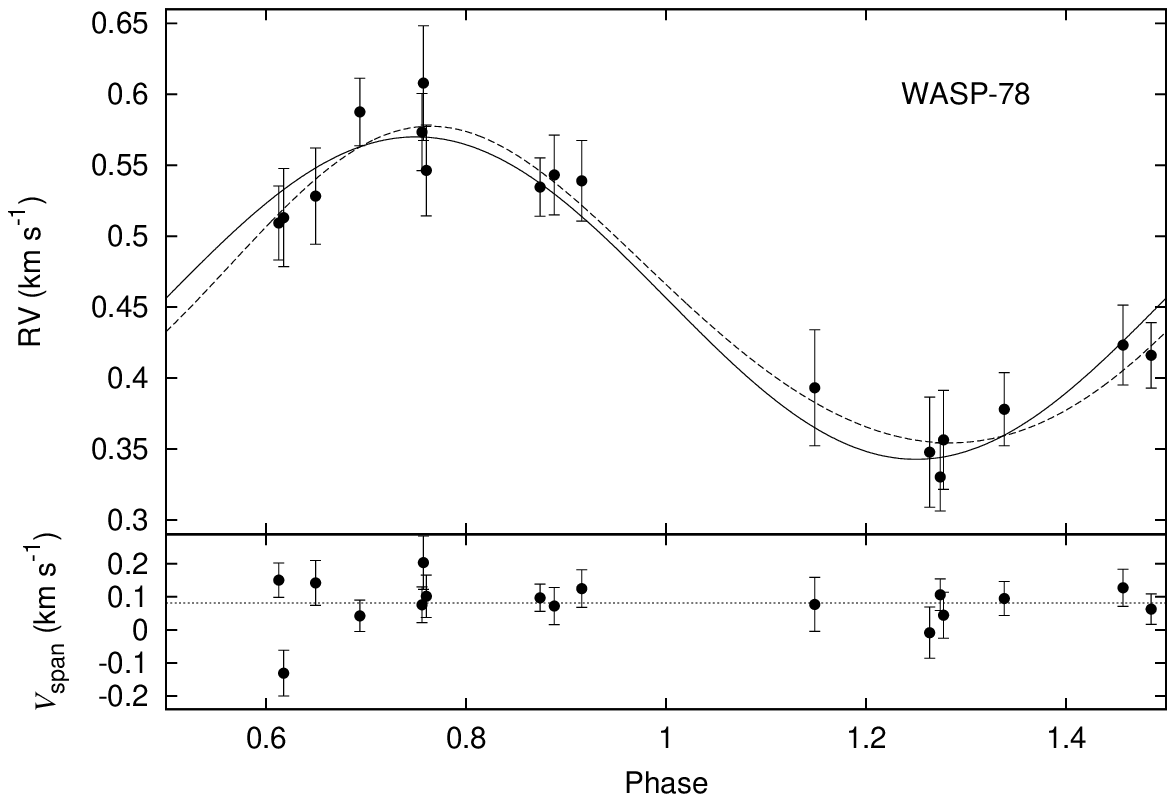}
\includegraphics[width=\columnwidth]{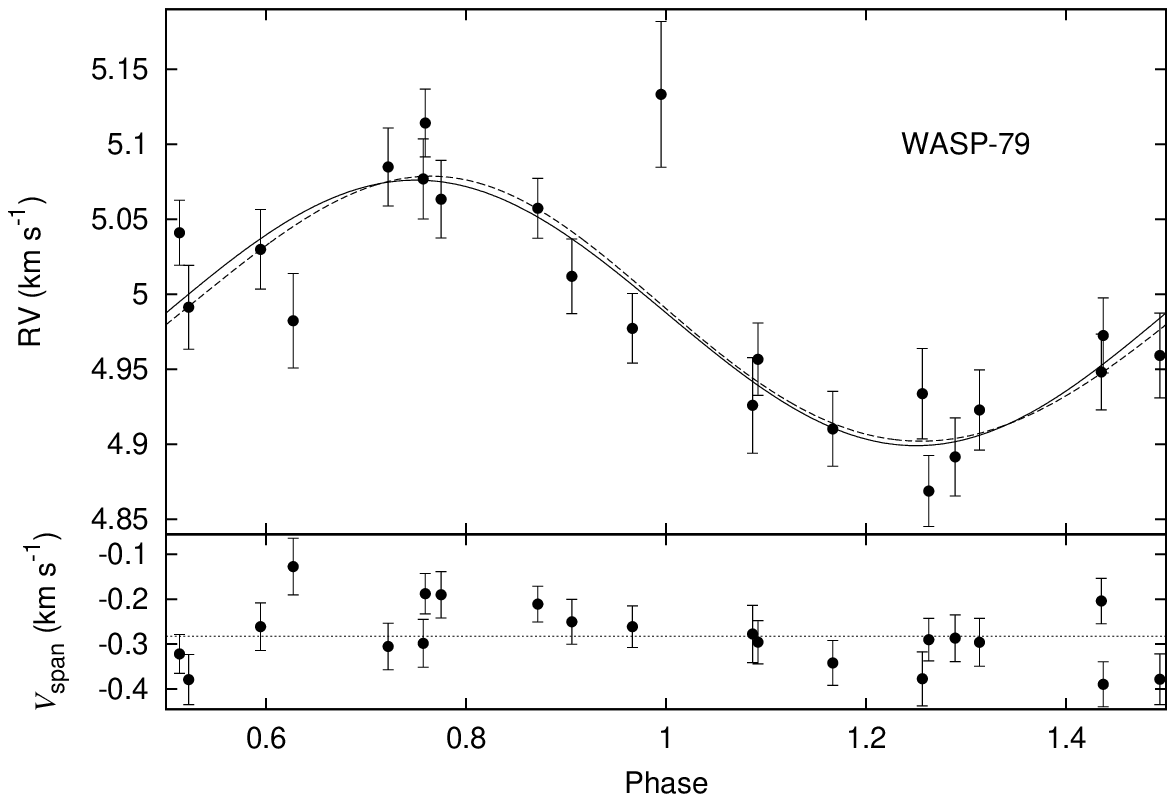}
\caption{Radial velocity variations and line bisectors ($V_{\rm span}$) of
WASP-78 and WASP-79 folded on the best-fitting circular orbit periods
(solid lines). The eccentric orbit solutions are shown as dashed-lines
(See Sect.~\ref{MCMC} for discussion). The bisector uncertainties of twice the
RV uncertainties have been adopted. The mean values of the bisectors are
indicated by the dotted lines. There is negligible correlation between $V_{\rm
span}$ and orbital phase.}
\label{WASP-RV}
\end{figure}

The amplitude of the RV variations and the absence of any
correlation with orbital phase of the line bisector spans ($V_{\rm span}$) in
Fig.~\ref{WASP-RV} indicates that it is highly improbable that the RV variations
are due to an unresolved eclipsing binary or chromospheric activity
\citep{2001A&A...379..279Q}.

In the case of WASP-79 one RV point (HJD 2455874.830894) was taken during
transit. This is affected by the Rossiter-McLaughlin effect and has been
excluded from the fitting process.

\section{TRAPPIST photometry}
\label{TRAPPIST_Phot}

Photometric observations of transits were obtained using the 60cm robotic
TRAPPIST Telescope
\citep{2011EPJWC..1106002G,2011A&A...533A..88G,2011Msngr.145....2J}. Since the
telescope has a German equatorial mount, there are short gaps in the lightcurves
at the times of culmination due to meridian flips, when the telescope is
reversed to the other side the pier.

\subsection{WASP-78b}
 
A total of three transits of WASP-78b were observed with TRAPPIST. The
observations were made through a `blue-blocking' filter with a transmittance
$>$90\% above 500nm. The exposure times were 15s and the telescope was
defocused, the mean FWHM being 5 pixels (pixel scale = 0.65\arcsec).

The first transit was observed on 2011 November 8 from 01h23 to 08h00 UTC. The
first $\approx$75\% of the run was slightly cloudy. There was a meridian flip at
HJD 2455873.748. The resulting lightcurve has 949 measurements. 

A second transit was observed on 2011 December 15 from 00h23 to 06h44 UTC.
Transparency was good this time.  A meridian flip occurred at HJD 2455910.646.
The resulting lightcurve has 918 measurements.

The third transit was observed on 2012 January 8 from 00h45 to 05h30 UTC.
However, only part of the transit was observed. Transparency was good. A
meridian flip occurred at HJD 2455934.579. The resulting lightcurve has 697
measurements.

\subsection{WASP-79b}

For WASP-79b only a partial transit was observed with TRAPPIST on 2011 September
26 from 07:33 to 09:52 UTC. It was obtained in $z'$ band. The exposure times
were 20s and the telescope was defocused by 300 focuser steps with the mean FWHM
being 5 pixels. The sky conditions were good. Due to a software problem, there
is a gap of a few minutes in the first half of the lightcurve. There was a
meridian flip at HJD 2455830.870. The resulting lightcurve has 244 measurements.

\section{Spectral Analysis}
\label{spectral}

The individual CORALIE spectra of WASP-78 and WASP-79 were co-added to produce
spectra with average S/N of around 50:1 and 90:1, respectively. The standard
pipeline reduction products were used in the analysis. The analysis was
performed using the methods given in \citet{2009A&A...496..259G}. The
km\,s$^{-1}$ and $H_\beta$ lines were used to determine the effective temperature
($T_{\rm eff}$). The surface gravity ($\log g$) was determined from the Ca~{\sc i} lines at
6162{\AA} and 6439{\AA} \citep{2010A&A...519A..51B}, along with the Na~{\sc i} D
lines. The Na D lines are affected by weak interstellar absorption. For
WASP-78 this is around 40m\AA, while for WASP-79 it is only around 10m\AA.
However, in both cases it is possible to fit the blue wings which are unaffected
by the interstellar lines. The elemental abundances were determined from
equivalent width measurements of several clean and unblended lines. A value for
microturbulence ($\xi_{\rm t}$) was determined from Fe~{\sc i} using the method of
\cite{1984A&A...134..189M}. The ionization balance between Fe~{\sc i} and
Fe~{\sc ii} and the null-dependence of abundance on excitation potential were
used as an additional the $T_{\rm eff}$ and $\log g$ diagnostics
\citep{2005MSAIS...8..130S}. The parameters obtained from the analysis are
listed in Table~\ref{stellar-params}. The quoted error estimates include that
given by the uncertainties in $T_{\rm eff}$, $\log g$ and $\xi_{\rm t}$, as well as the scatter
due to measurement and atomic data uncertainties. Also given are estimated
masses and radii using the \cite{2010A&ARv..18...67T} calibration. These values,
however, are not using in determining the planetary parameters, since {$T_{\rm eff}$}
and [Fe/H] are the only stellar input parameters (see Sect.~\ref{MCMC}).

\subsection{WASP-78}

There is no significant detection of lithium in the spectrum, with an equivalent
width upper limit of 4m\AA, corresponding to an abundance of $\log A$(Li) $<$
0.71 $\pm$ 0.12. This implies an age of at least several Gyr
\citep{2005A&A...442..615S}. A rotation rate of $P = 10.5 \pm 2.4$~d is implied
by the $v \sin i_{\star}$. This gives a gyrochronological age of $\sim
1.37^{+1.91}_{-0.78}$~Gyr using the \citet{2007ApJ...669.1167B} relation. 

\subsection{WASP-79}

There is lithium in the spectra, with an equivalent width of 8$\pm$1~m\AA,
corresponding to an abundance of $\log A$(Li) = 1.94 $\pm$ 0.07. The lithium
abundance of WASP-79 is an ineffective age constraint because of the high
effective temperature of this star. In addition, the star is close to the
lithium gap where depletion occurs in the 0.6~Gyr-old Hyades
\citep{1986ApJ...302L..49B,2004AJ....128.2435B}. A rotation rate of $P = 4.0 \pm
0.8$~d is implied by the $v \sin i_{\star}$. This gives a gyrochronological age of at
least $\sim 0.5$~Gyr using the \citet{2007ApJ...669.1167B} relation. However,
gyrochronology does not provide a reliable age constraint for such a hot
star.

\begin{table}[h]
\caption{Stellar parameters of WASP-78 and WASP-79}
\centering
\begin{tabular}{ccc} \hline\hline
Parameter     & WASP-78            & WASP-79 \\ \hline
Star          & TYC 5889-271-1     & CD-30 1812 \\
RA (J2000)    & 04h 15m 01.50s     & 04h 25m 29.01s \\
Dec (J2000)   &$-$22{\degr}06{\arcmin}59.0{\arcsec} &$-$30{\degr}36{\arcmin}01.5{\arcsec} \\
$V$ mag.      & 12.0               &10.1 \\
$T_{\rm eff}$ & 6100 $\pm$ 150 K   & 6600 $\pm$ 100 K \\
$\log g$      & 4.10 $\pm$ 0.20    & 4.20 $\pm$ 0.15   \\
$\xi_{\rm t}$ & 1.1 $\pm$ 0.2 km\,s$^{-1}$ & 1.3 $\pm$ 0.1 km\,s$^{-1}$ \\
$v_{\rm mac}$ & 3.5 $\pm$ 0.3 km\,s$^{-1}$ & 6.4 $\pm$ 0.3 km\,s$^{-1}$ \\
$v \sin i_{\star}$ & 7.2 $\pm$ 0.8 km\,s$^{-1}$ & 19.1 $\pm$ 0.7 km\,s$^{-1}$ \\
{[Fe/H]}      & $-$0.35 $\pm$ 0.14 &   +0.03 $\pm$ 0.10 \\
{[Si/H]}      & $-$0.25 $\pm$ 0.09 &   +0.06 $\pm$ 0.08 \\
{[Ca/H]}      & $-$0.14 $\pm$ 0.14 &   +0.25 $\pm$ 0.14 \\
{[Ti/H]}      & $-$0.02 $\pm$ 0.08 &       ...          \\
{[Cr/H]}      &          ...       &   +0.04 $\pm$ 0.19 \\
{[Ni/H]}      & $-$0.42 $\pm$ 0.10 & $-$0.06 $\pm$ 0.10 \\
$\log A$(Li)  & $<$ 0.71 $\pm$ 0.12 & 1.94 $\pm$ 0.07 \\
Mass          & 1.17 $\pm$ 0.13 $M_{\sun}$ & 1.38 $\pm$ 0.12 $M_{\sun}$\\
Radius        & 1.60 $\pm$ 0.41 $R_{\sun}$ & 1.53 $\pm$ 0.31 $R_{\sun}$  \\
Spectral Type &  F8   &   F5  \\
Distance      & 550 $\pm$ 120 pc & 240 $\pm$ 50 pc  \\ \hline
\end{tabular}
\label{stellar-params}

\tablefoot{Values of mactroturbulence ($v_{\rm mac}$) are based on the calibration by
\cite{2010MNRAS.405.1907B}. Mass and radius are estimated using the
\cite{2010A&ARv..18...67T} calibration. The spectral types are estimated from
$T_{\rm eff}$ using Table~B.1 in \cite{2008oasp.book.....G}.}

\end{table}

\section{Planetary system parameters}
\label{MCMC}

To determine the planetary and orbital parameters, a simultaneous fit the
CORALIE radial velocity measurements and both the WASP and TRAPPIST photometry
was performed, using the Markov Chain Monte Carlo (MCMC) technique. The details
of this process are described in \citet{2007MNRAS.380.1230C} and
\citet{2008MNRAS.385.1576P}. Four sets of solutions were used: with and without
the main-sequence mass-radius constraint (which imposes a Gaussian prior on the
stellar radius using a mass-radius relation $R_{\star} = M_{\star}^{0.8}$ with
$\sigma_R = 0.1 R$.) for both circular and floating eccentricity orbits.
Limb-darkening uses the four-coefficient model of \cite{2000A&A...363.1081C} and
vary with any change in {$T_{\rm eff}$} during the MCMC fitting process. For WASP data
and TRAPPIST with blue-blocking filter the $R$-band is used. The final values of
the limb-darkening coefficients are noted in Tables~\ref{wasp78b-mcmc} and
\ref{wasp79b-mcmc}.

A value of the stellar mass determined as part of the MCMC process, uses
the empirical calibration of \citet{2011MNRAS.417.2166S} in which mass is
estimated as a function of $T_{\rm eff}$, [Fe/H] and stellar
density ($\rho_{\star}$). The uncertainty in the derived stellar mass is dominated by the
uncertainties in the spectroscopic values of $T_{\rm eff}$ and [Fe/H]. At each
step in the Markov chain, these quantities are given random gaussian
perturbations, and controlled by priors assuming gaussian random errors in the
spectroscopic values. The stellar density is  derived at each step in the chain
from the scaled stellar radius $R_{\star}/a$ and the impact parameter ($b$). The
uncertainty in the stellar mass is computed directly from the posterior
probability distribution, and takes the uncertainties in $T_{\rm eff}$, [Fe/H]
and $\rho_{\star}$ fully into account. The posterior probability distribution
for the stellar radius follows from the mass and density values at each step in
the chain.

\subsection{WASP-78b}

\begin{figure}[h]
\includegraphics[width=\columnwidth]{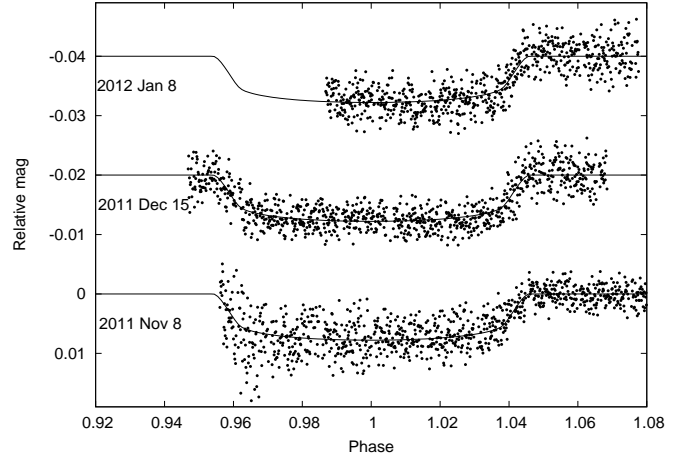}
\caption{TRAPPIST transits of WASP-78b. The solid-line is the
model fit presented in Sect.~\ref{MCMC}.
The lightcurves are offset for clarity.}
\label{phot-TRAPPIST-78} 
\end{figure}

The spectroscopic analysis suggests that WASP-78 might be slightly evolved. In
addition, we have three TRAPPIST lightcurves, which allows for the main-sequence
constraint to be removed (Fig.~\ref{phot-TRAPPIST-78}). The rms of the fits to
the TRAPPIST lightcurves is 2.33 {mmag.}, which is no better than that obtained
with the main-sequence constraint applied. However, the main-sequence constraint
forces the fitting process to a higher stellar mass (2.0~$M_{\sun}$), higher
$T_{\rm eff}$ (6640~K) and low impact parameter (0.06), whilst still finding a large
stellar radius (2.3~$R_{\sun}$) and, hence, planetary radius (1.75~$R_{\rm
Jup}$). Such a high values stellar mass and $T_{\rm eff}$ are inconsistent with the
spectroscopic analysis of the host star. Furthermore, fixing $T_{\rm eff}$
and [Fe/H] to their spectroscopic values yields a slightly worse
main-sequence-constrained fit to the photometry.

With the eccentricity floating, a value of $e = 0.078^{+0.054}_{-0.049}$ is
found. This fit has a $\chi^2 = 6.04$, compared to $\chi^2 = 7.92$ for a
circular orbit. Using the \citet[Eq. 27]{1971AJ.....76..544L} F-test shows that
there is a 17.2\% probability that the improvement could have arisen by chance.
This is too high to confidently claim detection of eccentricity. Thus, we
present the system parameters for the circular-orbit solutions
(Table~\ref{wasp78b-mcmc}), but note that the 3-$\sigma$ upper limit on
eccentricity is 0.24. \cite{2012MNRAS.422.1988A} presented a discussion on the
justification of adoption of circular orbit in these cases. Further RV
measurements, or occultation photometry, are required to constrain the possible
eccentricity of the system.

\begin{table}
\caption{System parameters for WASP-78b.}
\label{wasp78b-mcmc}
\centering
\begin{tabular}{ll} \hline\hline
Parameter & Value  \\ \hline
Transit epoch (HJD, UTC), $T_0$ &
2455882.35878  $\pm$ 0.00053
\\[+1mm]
Orbital period, $P$  &
2.17517632 $\pm$ 0.0000047 d
\\[+1mm]
Transit duration, $T_{14}$ &
0.1953 $\pm$ 0.0013 d
\\[+1mm]
Transit depth, $(R_{\rm P}/R_{*})^{2}$ &
0.0063 $\pm$ 0.0002
\\[+1mm]
Impact parameter, $b$ &
0.417 $^{+0.079}_{-0.129}$
\\[+1mm]
Stellar reflex velocity, $K_1$ &
0.1136 $\pm$ 0.0096 km\,s$^{-1}$
\\[+1mm]
Centre-of-mass velocity at time $T_0$, $\gamma$ &
0.4564 $\pm$ 0.0020 km\,s$^{-1}$
\\[+1mm]
Orbital separation, $a$ &
0.0362 $\pm$ 0.0008 AU
\\[+1mm]
Orbital inclination, $i$ &
83.2 $^{+2.3}_{-1.6}$ {\degr}
\\[+1mm]
Orbital eccentricity, $e$ &
0.0 (assumed)
\\[+1mm]
Stellar density, $\rho_{\star}$ &
0.125 $\pm$ 0.018 $\rho_{\sun}$
\\[+1mm]
Stellar mass, $M_{\star}$ &
1.33 $\pm$ 0.09 $M_{\sun}$
\\[+1mm]
Stellar radius, $R_{\star}$ &
2.20 $\pm$ 0.12 $R_{\sun}$
\\[+1mm]
Stellar surface gravity, $\log g_{\star}$ &
3.88 $\pm$ 0.04
\\[+1mm]
Planet mass, $M_{\rm P}$ &
0.89 $\pm$ 0.08 $M_{\rm Jup}$
\\[+1mm]
Planet radius, $R_{\rm P}$ &
1.70 $\pm$ 0.11 $R_{\rm Jup}$
\\[+1mm]
Planet surface gravity, $\log g_{\rm P}$ &
2.84 $\pm$ 0.06
\\[+1mm]
Planet density, $\rho_{\rm P}$ &
0.18 $\pm$ 0.04 $\rho_{\rm Jup}$
\\[+1mm]
Planet equilibrium temperature, $T_{\rm P}$ &
2350 $\pm$ 80 K
\\[+1mm]
\hline
\end{tabular}
\tablefoot{The planet equilibrium temperature, $T_{\rm P}$, assumes a Bond
albedo of $A=0$ and even redistribution of heat around the planet.
Limb-darkening coefficients were $a1$ = 0.388, $a2$ = 0.609, $a3$ = -0.395
and $a4$ = 0.075}
\end{table}

\subsection{WASP-79b}

With the eccentricity floating, a value of $e = 0.035^{+0.044}_{-0.024}$ is
found with a $\chi^2 = 20.9$, which is not significantly lower than $\chi^2 =
21.0$ for a circular orbit. Hence, there is no evidence for any orbital
eccentricity in the current RV data. Hence, we adopt a circular orbit for
WASP-79b.

\begin{figure}[h]
\includegraphics[width=\columnwidth]{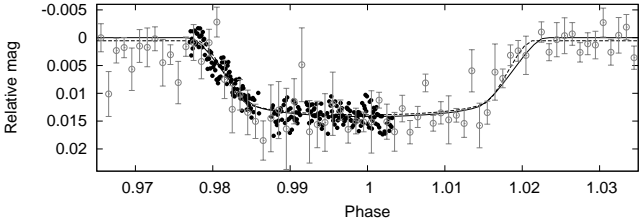}
\caption{TRAPPIST partial transit of WASP-79b. The solid-line is the non-main-sequence
constrained solution. The dashed-line is that with the main-sequence
constraint applied. The offset between the two fits reflects the zero-point
rescaling undertaken during the MCMC fits. The WASP photometry is shown as grey
circles with errorbars giving the standard deviation within the 0.001 phase
bins.}
\label{phot-TRAPPIST-79} 
\end{figure}

There is only a single partial TRAPPIST light curve and there are insufficient
points before ingress to reliably determine the out-of-transit level
(Fig.~\ref{phot-TRAPPIST-79}). The non-main-sequence constrained solution gives
an rms of 1.62 {mmag.} for the TRAPPIST photometry alone, compared to a slightly
larger value of 1.69 {mmag.} for that with the main-sequence constraint applied.
In addition, the main-sequence constraint has forced the fitting to a slightly
higher $T_{\rm eff}$ (6660~K) and [Fe/H] (+0.24) than found from the
spectroscopy. Fixing $T_{\rm eff}$ and [Fe/H] to their spectroscopic values
produces a slightly worse fit. The WASP photometry shows the depth is well
determined, but is unable help in constraining the shape and duration of the
ingress and egress. Thus, we present both the main-sequence and
non-main-sequence constrained circular orbit solutions in
Table~\ref{wasp79b-mcmc} and discuss them further in Sect.~\ref{Discussion}.

\begin{table*}
\caption{System parameters for WASP-79b. Two solutions are presented: column 2
gives that using the main-sequence constraint, while column 3 give that without
this constraint.}
\label{wasp79b-mcmc}
\centering
\begin{tabular}{lll} \hline\hline
Parameter & Value (ms) & Value (no-ms)\\ \hline
Transit epoch (HJD, UTC), $T_0$ &
2455545.23479 $\pm$ 0.00125&
2455545.23530 $\pm$ 0.00150
\\[+1mm]
Orbital period, $P$  &
3.6623817 $\pm$ 0.0000051 d &
3.6623866 $\pm$ 0.0000085 d
\\[+1mm]
Transit duration, $T_{14}$ &
0.1563 $\pm$ 0.0031 d &
0.1661 $\pm$ 0.0037 d 
\\[+1mm]
Transit depth, $(R_{\rm P}/R_{*})^{2}$ &
0.01148 $\pm$ 0.00051 &
0.01268 $\pm$ 0.00063
\\[+1mm]
Impact parameter, $b$ &
0.570 $\pm$ 0.052 &
0.706 $\pm$ 0.031
\\[+1mm]
Stellar reflex velocity, $K_1$ &
0.0882 $\pm$ 0.0078 km\,s$^{-1}$ &
0.0885 $\pm$ 0.0077 km\,s$^{-1}$
\\[+1mm]
Centre-of-mass velocity at time $T_0$, $\gamma$ &
4.9875 $\pm$ 0.0004 km\,s$^{-1}$ &
4.9875 $\pm$ 0.0004 km\,s$^{-1}$
\\[+1mm]
Orbital separation, $a$ &
0.0539 $\pm$ 0.0009 AU &
0.0535 $\pm$ 0.0008 AU
\\[+1mm]
Orbital inclination, $i$ &
85.4 $\pm$ 0.6 {\degr} &
83.3 $\pm$ 0.5 {\degr}
\\[+1mm]
Orbital eccentricity, $e$ &
0.0 (assumed) &
0.0 (assumed)
\\[+1mm]
Stellar density, $\rho_{\star}$ &
0.36 $\pm$ 0.04 $\rho_{\sun}$ &
0.22 $\pm$ 0.03 $\rho_{\sun}$
\\[+1mm]
Stellar mass, $M_{\star}$ &
1.56 $\pm$ 0.09 $M_{\sun}$ &
1.52 $\pm$ 0.07 $M_{\sun}$
\\[+1mm]
Stellar radius, $R_{\star}$ &
1.64 $\pm$ 0.08 $R_{\sun}$ &
1.91 $\pm$ 0.09 $R_{\sun}$
\\[+1mm]
Stellar surface gravity, $\log g_{\star}$ &
4.20 $\pm$ 0.03 &
4.06 $\pm$ 0.03
\\[+1mm]
Planet mass, $M_{\rm P}$ &
0.90 $\pm$ 0.09 $M_{\rm Jup}$ &
0.90 $\pm$ 0.08 $M_{\rm Jup}$
\\[+1mm]
Planet radius, $R_{\rm P}$ &
1.70 $\pm$ 0.11 $R_{\rm Jup}$ &
2.09 $\pm$ 0.14 $R_{\rm Jup}$
\\[+1mm]
Planet surface gravity, $\log g_{\rm P}$ &
2.85 $\pm$ 0.06 &
2.67 $\pm$ 0.06
\\[+1mm]
Planet density, $\rho_{\rm P}$ &
0.18 $\pm$ 0.04 $\rho_{\rm Jup}$ &
0.10 $\pm$ 0.02 $\rho_{\rm Jup}$
\\[+1mm]
Planet equilibrium temperature, $T_{\rm P}$ &
1770 $\pm$ 50 K &
1900 $\pm$ 50 K
\\[+1mm]
\hline
\end{tabular}
\tablefoot{The planet equilibrium temperature, $T_{\rm P}$, assumes a Bond
albedo of $A=0$ and even redistribution of heat around the planet.
Limb-darkening coefficients for the ms solution were for WASP $a1$ = 0.441, $a2$ = 0.599, $a3$ =
-0.514 and $a4$ = 0.150 and TRAPPIST ($z'$) $a1$ = 0.517, $a2$ = 0.202, $a3$ =
-0.181 and $a4$ = 0.025, while for the non-ms solution they were for 
WASP $a1$ = 0.447, $a2$ = 0.566, $a3$ =
-0.459 and $a4$ = 0.126 and TRAPPIST ($z'$) $a1$ = 0.523, $a2$ = 0.172, $a3$ =
-0.133 and $a4$ = 0.004.}
\end{table*}

\section{Discussion}
\label{Discussion}

\begin{figure}[h]
\includegraphics[width=\columnwidth]{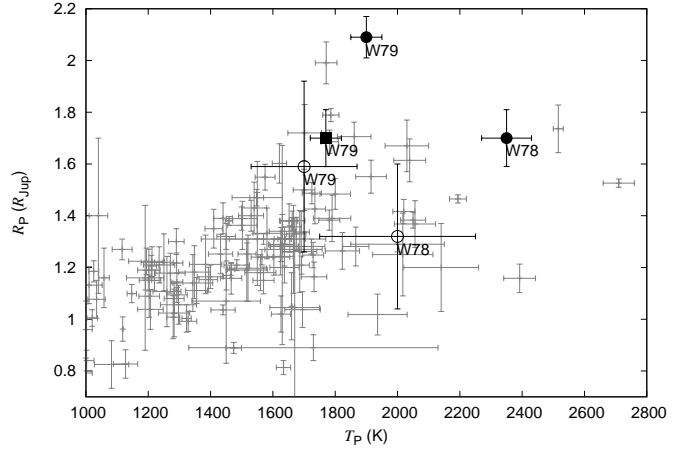}
\caption{Location of WASP-78b and WASP-79b in the exoplanet
equilibrium temperature--radius diagram. The filled-circles are the non-main
sequence constrained parameters, while the filled-square gives the main-sequence
constrained parameters for WASP-79b. The open circles shows the values
obtained using the stellar radii inferred from the spectroscopic $\log g$ values
(See text for discussion).
Values for other exoplanets were taken from TEPCAT \citep{2011MNRAS.417.2166S}.}
\label{pl-t-r} 
\end{figure}

WASP-78b and WASP-79b are two Jupiter-mass exoplanets transiting the F-type host
stars. They occupy the region in parameter space of large radius, low density,
exoplanets. The high planetary equilibrium temperature ($T_{\rm P} = 2350 \pm
80$~K) for WASP-78b, makes this one of the hottest exoplanets currently known
(Fig.~\ref{pl-t-r}), exceeded only by WASP-33b \citep{2010MNRAS.407..507C},
WASP-12b \citep{2009ApJ...693.1920H} and WASP-18b \citep{2009Natur.460.1098H}.
The mass, radius and density of WASP-78b, are similar to those of HAT-P-32b
\citep[their circular orbit solution]{2011ApJ...742...59H}, OGLE-TR-10b
\citep{2005ApJ...624..372K} and TrES-4b \citep{2011AJ....141..179C}. Although,
similar in radius, WASP-12b is a much denser, more massive planet
\citep{2009ApJ...693.1920H}. The large planetary radius of WASP-79b could even
exceed that of WASP-17b \citep{2010ApJ...709..159A,2011MNRAS.416.2108A}.
However, due to its higher mass, WASP-79b's density is somewhat higher. The
main-sequence constrained solution for WASP-79b is, on the other hand, somewhat
similar to circular orbit parameters for HAT-P-33b \citep{2011ApJ...742...59H}.

The planetary radius and equilibrium temperature are dependant on the inferred
stellar radius. Referring back to the spectral analysis results
(Table~\ref{stellar-params}) and using the stellar radii obtained from the
\cite{2010A&ARv..18...67T} relationships, yields planetary radii of $R_{\rm P}$
= 1.32 $\pm$ 0.28 $R_{\rm Jup}$ and 1.59 $\pm$ 0.33 $R_{\rm Jup}$ for WASP-78b
and WASP-79b, respectively. Both are smaller than those obtained from the
transit lightcurves, especially WASP-78b which is slightly over 1-$\sigma$ away
from that obtained from the transit analysis. The equilibrium temperatures for
both planets would fall to 2000 $\pm$ 250 K and 1700 $\pm$ 170 K, respectively.
Of course, these results are dependent on the rather imprecise spectroscopic
$\log g$ values, where a relatively small change can have a dramatic effect on
the inferred stellar radius.

\begin{figure}[h]
\includegraphics[width=\columnwidth]{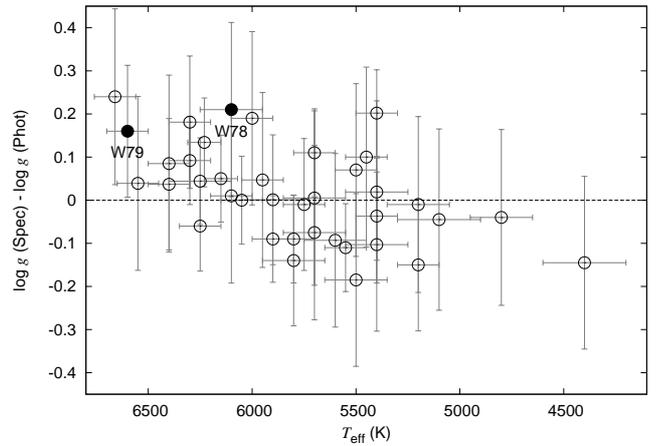}
\caption{Difference between spectroscopic $\log g$ determined from CORALIE
spectra for WASP hosts stars and that obtained from transit photometry, as a
function of stellar effective temperature ($T_{\rm eff}$).}
\label{logg} 
\end{figure}

The surface gravities implied by the transit fits for both WASP-78 and WASP-79
are lower than those found from the spectroscopic analyses. Spectroscopic
$\log g$ determinations have a typical uncertainty of 0.2~dex
\citep{2005MSAIS...8..130S} and are less precise than those determined from
planetary transit modelling. In order to evaluate whether there are any
systematic differences, we have compared the spectroscopic $\log g$ values
obtained in previous CORALIE analyses of WASP planet-host stars with the values
determined using transit photometry (Fig.~\ref{logg}). There is a large scatter
in the differences, due to uncertainties in the spectroscopic values, but there
does not appear to be any significant systematic difference. There is a hint
that the difference may increase for the hottest stars, which could be due to
systematic N-LTE effects in the spectroscopic values. In fact,
\cite{2012MNRAS.423..122B} found a similar result from their spectroscopic
analyses of stars with asteroseismic $\log g$ values obtained using {\it Kepler}.

The relatively large difference in $\log g$ from spectroscopy and transit
photometry for WASP-78 is probably due to the low S/N of the CORALIE spectrum,
where the effects of scattered light and uncertainties in continuum location can
result in slight systematics. The $T_{\rm eff}$ values obtained from CORALIE spectra
have been found to be in agreement with those from the infrared flux method
\citep{2011MNRAS.418.1039M}, suggesting that a change of more than 100~K is
unlikely. However, \citet{DOYLE} found that {[Fe/H]} obtained using CORALIE
spectra is, on average, 0.08 $\pm$ 0.05~dex lower than found using higher S/N
HARPS spectra. Hence, a modest decrease in $T_{\rm eff}$ and/or increase in {[Fe/H]}
could reduce, or even eliminate, the $\log g$ discrepancy. A high S/N spectrum is
required to improve the precision of the spectroscopic parameters. In the case
of WASP-79, the spectroscopic $\log g$ value could be too high by of the order
0.1~dex due to N-LTE effects. A value of $\log g$ = 4.1 $\pm$ 0.15, would give a
stellar radius of 1.76 $\pm$ 0.37, and a planetary radius of 1.83 $\pm$
0.39~$R_{\rm Jup}$ for WASP-79b. This is now close to the non-main sequence
constrained $\log g$ of 4.06$\pm$0.03, suggesting that the system parameters
are closer to this solution. However, given that the photometry cannot constrain
the transit shape, the planetary radius may not be as extreme as 2.09 $\pm$ 0.14
$R_{\rm Jup}$. Thus, we conclude that the planetary radius is at least 1.70
$\pm$ 0.11 $R_{\rm Jup}$, but possibly somewhat larger. Further transit
photometry is required to confirm the planetary parameters.

The empirical relationship for Jupiter-mass exoplanets of
\citet[Eq.~9]{2012A&A...540A..99E}, predicts planetary radii of 1.78 and 1.56
(1.66 for non-main-sequence solution) for WASP-78b and WASP-79b, respectively.
These are in good agreement with that found for WASP-78b, but somewhat smaller
than found for WASP-79b (in either of the two cases presented).
\cite{2012A&A...540A..99E} noted that their relationship also significantly
underestimated the radius of HAT-P-32b, suggesting that tidal heating could
explain the extra inflation, over and above that from stellar irradiation.
However, \cite{2012A&A...540A..99E} used the eccentric solution from
\cite{2011ApJ...742...59H} which \cite{2012MNRAS.422.1988A} concluded this is
likely to be incorrect and the circular solution should be adopted. This makes
HAT-P-32b slightly less bloated, but still larger than given by the
empirical relationship. The radial velocities presented here do
not preclude the presence of modest eccentricity in either or both of the
WASP-78 and WASP-79 systems. Further observations of these two systems are
required to fully constrain their properties.

\section*{Acknowledgements}

WASP-South is hosted by the South African Astronomical Observatory and their
support and assistance is gratefully acknowledged. TRAPPIST is a project funded
by the Belgian Fund for Scientific Research (Fond National de la Recherche
Scientifique, F.R.S-FNRS) under grant FRFC 2.5.594.09.F, with the participation
of the Swiss National Science Fundation (SNF). M. Gillon and E. Jehin are FNRS
Research Associates. M. Gillon also acknowledges support from the Belgian
Science Policy Office in the form of a Return Grant.  This research has made use
of the SIMBAD database, operated at CDS, Strasbourg, France.

\bibliographystyle{aa}

\begin{thebibliography}{}

\bibitem[Anderson et al.(2010)]{2010ApJ...709..159A} Anderson, D.R., Hellier,
C., Gillon, M., et al. 2010, \apj, 709, 159
	
\bibitem[Anderson et al.(2011)]{2011MNRAS.416.2108A} Anderson, D.R.,
Smith, A.M.S., Lanotte, A.A., et al. 2011, \mnras, 416, 2108

\bibitem[Anderson et al.(2012)]{2012MNRAS.422.1988A} Anderson, D.~R., 
Collier Cameron, A., Gillon, M., et al.\ 2012, \mnras, 422, 1988 

\bibitem[Arras \& Socrates(2010)]{2010ApJ...714....1A} Arras, P., \& Socrates,
A.\ 2010, \apj, 714, 1

\bibitem[Baranne et al.(1996)]{1996A&AS..119..373B} Baranne, A., Queloz, D.,
Mayor, M., et al. 1996, \aaps, 119, 373

\bibitem[Barnes(2007)]{2007ApJ...669.1167B} Barnes, S.A. 2007, \apj, 669, 1167

\bibitem[Batygin \& Stevenson(2010)]{2010ApJ...714L.238B} Batygin, K., \&
Stevenson, D.~J.\ 2010, \apjl, 714, L238 

\bibitem[Boesgaard \& Tripicco(1986)]{1986ApJ...302L..49B} Boesgaard, A.~M., \&
Tripicco, M.~J.\ 1986, \apjl, 302, L49 

\bibitem[B{\"o}hm-Vitense(2004)]{2004AJ....128.2435B} B{\"o}hm-Vitense, E.\ 
2004, \aj, 128, 2435 

\bibitem[Bruntt et al.(2010a)]{2010A&A...519A..51B} Bruntt, H., Deleuil, M.,
Fridlund, M., et al. 2010, \aap, 519, A51

\bibitem[Bruntt et al.(2010)]{2010MNRAS.405.1907B} Bruntt H., Bedding T.R.,
Quirion P.-O., et al., 2010, \mnras, 405, 1907

\bibitem[Bruntt et al.(2012)]{2012MNRAS.423..122B} Bruntt, H., Basu, S., 
Smalley, B., et al.\ 2012, \mnras, 423, 122

\bibitem[Chan et al.(2011)]{2011AJ....141..179C} Chan, T., Ingemyr, M., 
Winn, J.~N., et al.\ 2011, \aj, 141, 179 

\bibitem[Charbonneau et al.(2000)]{2000ApJ...529L..45C} Charbonneau, D., 
Brown, T.~M., Latham, D.~W., \& Mayor, M.\ 2000, \apjl, 529, L45

\bibitem[Charbonneau et al.(2005)]{2005ApJ...626..523C} Charbonneau, D., Allen,
L.E., Megeath, S.T., Torres, G., \& Alonso, R. 2005, \apj, 626, 523

\bibitem[Claret(2000)]{2000A&A...363.1081C} Claret, A., 2000, \aap, 363, 1081

\bibitem[Collier Cameron et al.(2006)]{2006MNRAS.373..799C} Collier Cameron, A.,
Pollacco, D.L., Street, R.A., et al. 2006, \mnras, 373, 799

\bibitem[Collier Cameron et al.(2007)]{2007MNRAS.380.1230C} Collier Cameron, A.,
Wilson, D.M., West, R.G., et al. 2007, \mnras, 380, 1230

\bibitem[Collier Cameron et al.(2010)]{2010MNRAS.407..507C} Collier 
Cameron, A., Guenther, E., Smalley, B., et al.\ 2010, \mnras, 407, 507 

\bibitem[Enoch et al.(2012)]{2012A&A...540A..99E} Enoch, B., Collier Cameron, A.,
\& Horne, K.\ 2012, \aap, 540, A99 


\bibitem[Doyle et al.(2012)]{DOYLE} Doyle, A.~P., Smalley, B., Maxted, P.~F.~L.,
et al.\ 2012, \mnras, submitted.

\bibitem[Fortney et al.(2007)]{2007ApJ...659.1661F} Fortney, J.~J., Marley, 
M.~S., \& Barnes, J.~W.\ 2007, \apj, 659, 1661 

\bibitem[Fortney et al.(2011)]{2011ApJS..197....9F} Fortney, J.~J., Demory, 
B.-O., D{\'e}sert, J.-M., et al.\ 2011, \apjs, 197, 9 

\bibitem[Fortney  \& Nettelmann(2010)]{2010SSRv..152..423F} Fortney, J.~J., \&
Nettelmann, N.\ 2010, \ssr, 152, 423  

\bibitem[Gillon et al.(2009)]{2009A&A...496..259G} Gillon, M., Smalley, B.,
Hebb, L., et al. 2009, \aap, 496, 259

\bibitem[Gillon et al.(2011a)]{2011EPJWC..1106002G} Gillon, M., Jehin, E., 
Magain, P., et al.\ 2011, Detection and Dynamics of Transiting Exoplanets, 
St.~Michel l'Observatoire, France, Edited by F.~Bouchy; R.~D{\'{\i}}az; 
C.~Moutou; EPJ Web of Conferences, Volume 11, id.06002, 11, 6002 

\bibitem[Gillon et al.(2011b)]{2011A&A...533A..88G} Gillon, M., Doyle, A.~P.,
Lendl, M., et al.\ 2011, \aap, 533, A88 

\bibitem[Gray(2008)]{2008oasp.book.....G} Gray, D.F., 2008, The observation and analysis of
stellar photospheres, 3rd Edition (Cambridge University Press), p.~507.

\bibitem[Hartman et al.(2011)]{2011ApJ...742...59H} Hartman, J.~D., Bakos, 
G.~{\'A}., Torres, G., et al.\ 2011, \apj, 742, 59 

\bibitem[Hebb et al.(2009)]{2009ApJ...693.1920H} Hebb, L., Collier-Cameron, 
A., Loeillet, B., et al.\ 2009, \apj, 693, 1920 

\bibitem[Hellier et al.(2009)]{2009Natur.460.1098H} Hellier, C., Anderson, 
D.~R., Collier Cameron, A., et al.\ 2009, \nat, 460, 1098 

\bibitem[Jackson \& Stoy(1954)]{1954AnCap..17.....J} Jackson, J., \& Stoy,
R.~H.\ 1954, Annals of the Cape Observatory, 17A 

\bibitem[Jehin et al.(2011)]{2011Msngr.145....2J} Jehin, E., Gillon, M., 
Queloz, D., et al.\ 2011, The Messenger, 145, 2 

\bibitem[Jones et al.(2009)]{2009MNRAS.399..683J} Jones, D.H., Read, M.A.,
Saunders, W., et al. 2009, \mnras, 399, 683

\bibitem[Konacki et al.(2005)]{2005ApJ...624..372K} Konacki, M., Torres, 
G., Sasselov, D.~D., \& Jha, S.\ 2005, \apj, 624, 372 

\bibitem[Latham et al.(2010)]{2010ApJ...713L.140L} Latham, D.~W., Borucki, 
W.~J., Koch, D.~G., et al.\ 2010, \apjl, 713, L140 

\bibitem[Lucy \& Sweeney(1971)]{1971AJ.....76..544L} Lucy, L.B., \& Sweeney,
M.A. 1971, \aj, 76, 544

\bibitem[Magain(1984)]{1984A&A...134..189M} Magain, P. 1984, \aap, 134, 189

\bibitem[Maxted et al.(2011)]{2011MNRAS.418.1039M} Maxted, P.~F.~L., Koen, 
C., \& Smalley, B.\ 2011, \mnras, 418, 1039 

\bibitem[Mayor \& Queloz(1995)]{1995Natur.378..355M} Mayor, M., \& Queloz, D.
1995, \nat, 378, 355

\bibitem[Pepe et al.(2002)]{2002A&A...388..632P} Pepe, F., Mayor, M.,
Galland, F., et al. 2002, \aap, 388, 632

\bibitem[Perna et al.(2010)]{2010ApJ...724..313P} Perna, R., Menou, K., 
\& Rauscher, E.\ 2010, \apj, 724, 313

\bibitem[Pollacco et al.(2006)]{2006PASP..118.1407P} Pollacco, D.L.,
Skillen, I., Collier Cameron, A., et al. 2006, \pasp, 118, 1407

\bibitem[Pollacco et al.(2008)]{2008MNRAS.385.1576P} Pollacco, D.L.,
Skillen, I., Collier Cameron, A., et al. 2008, \mnras, 385, 1576

\bibitem[Queloz et al.(2000)]{2000A&A...354...99Q} Queloz, D., Mayor, M.,
Weber, L., et al. 2000, \aap, 354, 99

\bibitem[Queloz et al.(2001)]{2001A&A...379..279Q} Queloz, D., Henry, G.W.,
Sivan, J.P., et al. 2001, \aap, 379, 279

\bibitem[Sestito \& Randich(2005)]{2005A&A...442..615S} Sestito, P., \& Randich,
S., 2005 \aap, 442, 615

\bibitem[Smalley(2005)]{2005MSAIS...8..130S} Smalley, B. 2005, Mem. Soc. Astron.
Ital. Suppl., 8, 130

\bibitem[Southworth(2009)]{2009MNRAS.394..272S} Southworth, J. 2009, \mnras,
394, 272
	
\bibitem[Southworth(2009)]{2011MNRAS.417.2166S} Southworth, J. 2011, \mnras,
417, 2166

\bibitem[Torres et al.(2010)]{2010A&ARv..18...67T}
Torres, G., Andersen, J. \& Gim\'enez, A, 2010a, \aapr, 18, 67

\bibitem[Winn(2009)]{2009IAUS..253...99W} Winn, J.N., 2009, IAU Symposium, 253,
 99

\end{thebibliography}

\end{document}